\documentclass[10pt,aps,prd,nofootinbib,reprint,superscriptaddress]{revtex4-1}

\usepackage[utf8]{inputenc}
\usepackage{amsmath}
\usepackage{graphicx}
\usepackage{hyperref}
\hypersetup{colorlinks=true,allcolors=[rgb]{1,0.56,0}}

\newcommand{\orcid}[1]{\href{https://orcid.org/#1}{#1}}
\newcommand{\ket}[1]{|#1\rangle}
\newcommand{\wt}[1]{\widetilde{#1}}

\begin{document}

\title{Neutrino Oscillations through the Earth's Core}

\author{Peter B.~Denton}
\thanks{\orcid{0000-0002-5209-872X}}
\email{pdenton@bnl.gov}
\affiliation{High Energy Theory Group, Physics Department, Brookhaven National Laboratory, Upton, NY 11973, USA}

\author{Rebekah Pestes}
\thanks{\orcid{0000-0002-9634-1664}}
\email{rebhawk8@vt.edu}
\affiliation{High Energy Theory Group, Physics Department, Brookhaven National Laboratory, Upton, NY 11973, USA}
\affiliation{Center for Neutrino Physics, Department of Physics, Virginia Tech, Blacksburg, VA 24061, USA}
\date{\today}

\begin{abstract}
Neutrinos have two properties that make them fairly unique from other known particles: extremely low cross sections and flavor changing oscillations.
With a good knowledge of the oscillation parameters soon in hand, it will become possible to detect low-energy atmospheric neutrinos sensitive to the forward elastic scattering off electrons in the Earth's core providing a measurement of the core properties and the matter effect itself.
As the dynamics of the Earth's core are complicated and in a difficult to probe environment, additional information from upcoming neutrino experiments will provide feedback into our knowledge of geophysics as well as useful information about exoplanet formation and various new physics scenarios including dark matter.
In addition, we can probe the existence of the matter effect in the Earth and constrain the non-standard neutrino interaction parameter $\epsilon_{ee}^\oplus$.
We show how DUNE's sensitivity to low-energy atmospheric neutrino oscillations can provide a novel constraint on the density and radius of the Earth's core at the 9\% level and the Earth's matter effect at the 5\% level.
Finally, we illuminate the physics behind low-energy atmospheric neutrino resonances in the Earth.
\end{abstract}

\date{\today}

\maketitle

\section{Introduction}
Even as we are still measuring the remaining known neutrino parameters, neutrinos provide a unique window into difficult to probe environments.
Because of their extremely small cross sections, neutrinos provide crucial information about the Sun \cite{Vinyoles:2016djt,BOREXINO:2020aww}, supernova \cite{Kamiokande-II:1987idp,Bionta:1987qt,Alekseev:1987ej,Moller:2018kpn,Migenda:2019xbm,Hyper-Kamiokande:2021frf}, nuclear processes within the Earth \cite{KamLAND:2013rgu,Borexino:2019gps,NAP12161}, nuclear reactors \cite{Bernstein:2019hix}, and the big bang \cite{Oldengott:2017fhy,Lancaster:2017ksf,Kreisch:2019yzn,Barenboim:2019tux,Mazumdar:2019tbm} making neutrinos an excellent means of extracting information from dense and difficult to access environments.

While their cross sections are quite low, at high energies (TeV-PeV), neutrinos begin to become absorbed in the Earth.
This effect has been used along with IceCube data to determine the density of the different layers of the Earth \cite{Donini:2018tsg}.

All of the above probes leverage the neutrino's small, but non-zero, cross section.
In addition to their small cross section, neutrinos also undergo flavor changing oscillations during propagation \cite{Pontecorvo:1957cp,Super-Kamiokande:1998kpq,SNO:2002tuh}.
During propagation in matter, neutrino oscillations are modified by the presence of electrons \cite{Wolfenstein:1977ue}.
This allows one, in principle, to probe the electron number density of dense materials \cite{Agarwalla:2012uj,Rott:2015kwa,Winter:2015zwx,Bourret:2017tkw,Bakhti:2020tcj} assuming either the oscillation parameters are measured in vacuum or measurements are made in different environments.

We propose measuring the core of the Earth using neutrinos produced in the atmosphere that then oscillate during propagation in the Earth as modified by the matter effect, and detected in the large DUNE LArTPC detectors.
In addition, we show how the Earth's matter effect can be measured with atmospheric neutrinos at DUNE, and in the appendix, we elucidate several features of resonant oscillations.

\section{Atmospheric Neutrino Oscillations}
\subsection{Overview}
Cosmic ray interactions in the atmosphere produce an abundant flux of neutrinos that has been measured over many decades of energy.
As these neutrinos propagate through the Earth to a detector, they oscillate and also experience the matter effect \cite{Wolfenstein:1977ue} from forward elastic scattering with electrons inside the Earth.

Neutrino propagation in matter is calculated by solving the Schr\"odinger equation,
\begin{equation}
i\frac{\partial}{\partial x}\ket{\nu}=H\ket{\nu}\,,
\end{equation}
where the Hamiltonian for neutrino oscillations in varying matter potential in the Earth in the flavor basis is
\begin{equation}
H=\frac1{2E}\left[U
\begin{pmatrix}
0\\&\Delta m^2_{21}\\&&\Delta m^2_{31}
\end{pmatrix}
U^\dagger
+
\begin{pmatrix}
a(x)\\&0\\&&0
\end{pmatrix}\right]\,.
\end{equation}
Here $U$ is the lepton mixing matrix \cite{Pontecorvo:1957cp,Maki:1962mu} and $a(x)=2\sqrt2G_FN_e(x)E$ is the matter potential.
The electron number density $N_e(x)$ varies with propagation distance and this is the primary effect that we are investigating.

For the atmospheric flux, we use \cite{Honda:2015fha} and take a conservative 40\% normalization uncertainty on each of the four production channels ($\nu_\mu,\nu_e,\bar\nu_\mu,\bar\nu_e$) separately and a spectral index uncertainty of $0.2$.

Neutrino oscillations in the core of the Earth\footnote{For an alternative discussion of neutrinos in the Earth's core, see e.g.~\cite{2012}.} experience a slight parametric resonance when crossing the core \cite{Akhmedov:1988kd,Krastev:1989ix}.
A parametric resonance occurs approximately when
\begin{align}
\left|\frac{\Delta\wt{m^2_{21}}_mL_m}{4E}\right|&=\frac\pi2(2k_m+1)\,,\\
\left|\frac{\Delta\wt{m^2_{21}}_cL_c}{4E}\right|&=\frac\pi2(2k_c+1)\,,
\end{align}
are both simultaneously satisfied for some integers $k_m,k_c$ where $\Delta\wt{m^2_{21}}_{m,c}$ is $\Delta m^2_{21}$ in the mantle or the core, respectively and $L_{m,c}$ is the distance traveled in the mantle up to the core and through the core, respectively.
The conditions for the parametric resonance are satisfied for a modest range of energies $\sim175-325$ MeV, but only a very narrow range of zenith angles centered at $\cos\theta_z\simeq-0.84$ with a width of $\Delta\cos\theta_z\lesssim0.001$ at $k_c=0$ and $k_m=1$.
The exact values differ somewhat for antineutrinos and the most relevant resonance occurs for $k_c=k_m=1$, but the narrowness of the $\cos\theta_z$ range remains similar.
Because of this, any effects from a parametric resonance will be washed out in a detector.
Atmospheric neutrinos in the Earth do experience an MSW \cite{Mikheyev:1985zog} resonant enhancement of the mixing angle, but it either occurs at energies below DUNE's sensitivity or has a negligible contribution to the flux.
See appendix \ref{sec:resonances} for more on resonances.

\subsection{The Earth's Core}
Our understanding of the inner structure of the Earth depends primarily on seismographic data.
Seismic wave propagation is affected by the composition, pressure, and temperature of the material, as well as the exact location and depth of the earthquake, which makes a precise determination somewhat difficult.
The inner core is believed to be solid while the outer core is believed to be liquid based on information coming from the propagation of S- and P-waves.
The dynamics of the core of the Earth are quite complicated, involving the Coriolis effect, Lorentz forces, a self-excited dynamo mechanism, and Alfven waves propagating on timescales of 3-4 years, and are related to the Earth's magnetic field \cite{Buffett2010,SchubertTreatise,Larmor,Glatzmaiers1995,Nataf2014,PhysRevE.83.066310,Gillet2010}, which undergoes flips and other significant modifications on difficult to predict timescales \cite{NOWACZYK201254}.
Moreover, there are connections between understanding the core of our planet and the cores of exoplanets \cite{Armstrong2020}.
In addition, it is speculated that the core may have modest eccentricity, although this is not yet fully understood \cite{doi:10.1080/08120090801888578,https://doi.org/10.1029/GL013i013p01545,https://doi.org/10.1029/94GL01600,tanaka2007possibility,song1998seismic,karato2000earth,ishii2002innermost,frost2021dynamic}.
Finally, the outer core's radius is composed of the so-called D'' layer between the core and the mantle, which is $\sim200$ km thick and whose properties and topology are fairly uncertain \cite{https://doi.org/10.1029/GL013i013p01497} and may be the location of the formation of hotspots within the mantle \cite{Geodesy1993}.
Further knowledge of the core of the Earth has broad interest in the context of understanding how the Earth formed, the nature of the Earth's magnetic field and its dynamics, as well as exoplanet formation.

\begin{figure}
\centering
\includegraphics[width=\columnwidth]{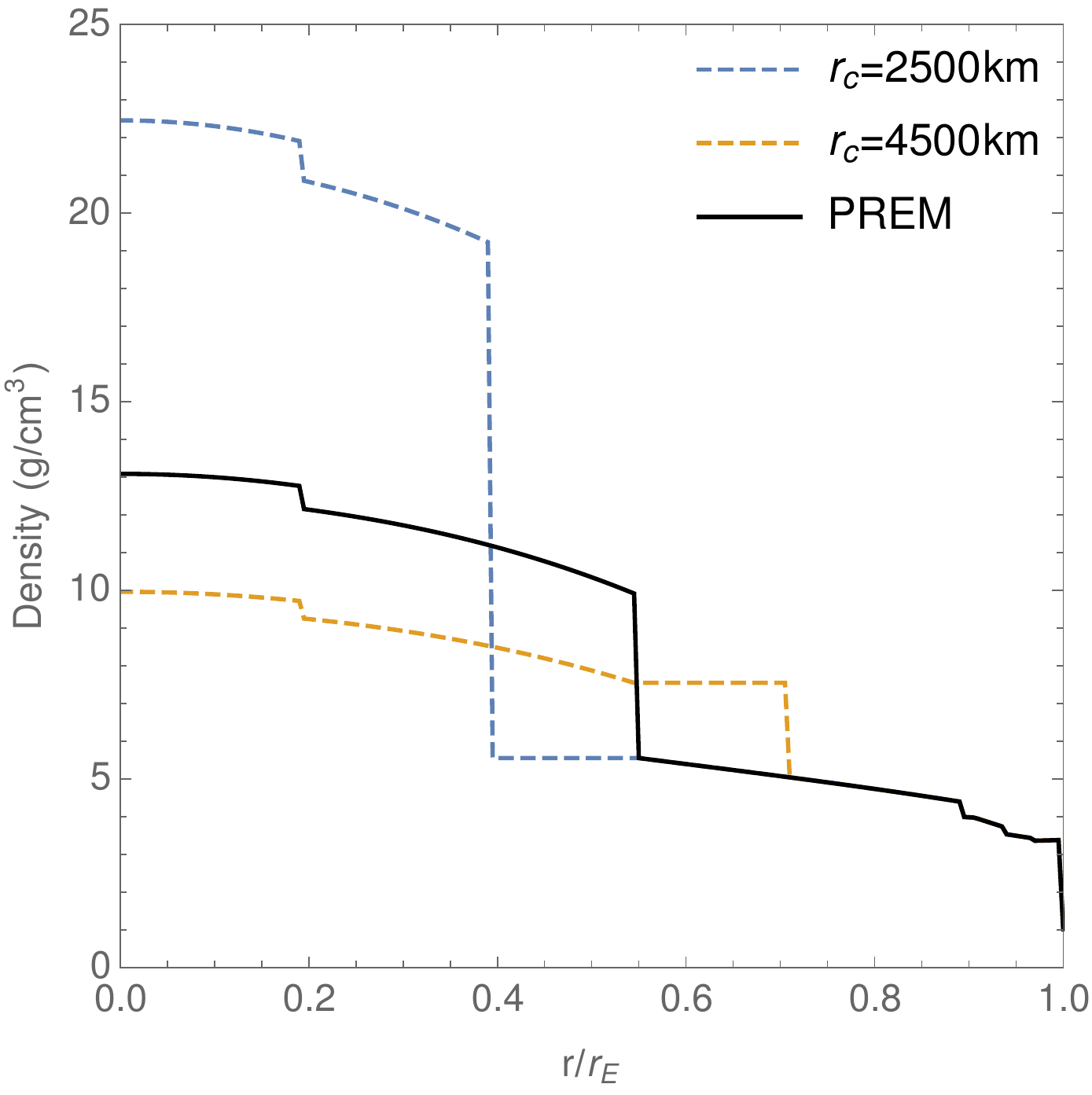}
\caption{The radial density profile of the Earth as a function of radius in units of $r_E=6371$ km as well as the density profiles for small and large core examples.}
\label{fig:prem variation}
\end{figure}

To probe the Earth's core with neutrinos, we vary both the core radius and the core density simultaneously, keeping the total mass of the Earth constant \cite{ParticleDataGroup:2020ssz}.
That is, given the density of the Earth as a function of its radius according to the PREM $\rho_{\rm PREM}^0(r)$, we vary the outer core radius $r_c$ from its fiducial value $r_c^0=3480$ km by
\begin{equation}
\rho(r)=
\begin{cases}
C\rho_{\rm PREM}^0(r)&r<r_c<r_c^0\\
\rho_{\rm PREM}^0(r_c^0)_+&r_c<r<r_c^0\\
\rho_{\rm PREM}^0(r)&r_c<r_c^0<r\\
\end{cases}
\end{equation}
for small core radii and
\begin{equation}
\rho(r)=
\begin{cases}
D\rho_{\rm PREM}^0(r)&r<r_c^0<r_c\\
D\rho_{\rm PREM}^0(r_c^0)_-&r_c^0<r<r_c\\
\rho_{\rm PREM}^0(r)&r_c^0<r_c<r\\
\end{cases}
\end{equation}
for large core radii, where $\rho_{\rm PREM}^0$ is the density of the Earth from the standard PREM model based on an analysis of seismographic data \cite{Dziewonski:1981xy}, $\rho_{\rm PREM}^0(r_c^0)_\pm$ is the limit of the density as $r\to r_c^0$ from above or below, respectively.
The scaling factors $C,D$ are given by
\begin{align}
C&=\frac{\int_0^{r_c^0}drr^2\rho_{\rm PREM}^0(r)-\frac{\rho(r_c^0)_+}3\left[(r_c^0)^3-r_c^3\right]}{\int_0^{r_c}drr^2\rho_{\rm PREM}^0(r)}\\
D&=\frac{\int_0^{r_c}drr^2\rho_{\rm PREM}^0(r)}{\int_0^{r_c^0}drr^2\rho_{\rm PREM}^0(r)+\frac{\rho(r_c^0)_-}3\left[r_c^3-(r_c^0)^3\right]}
\end{align}
This keeps the total mass of the Earth constant while changing the density and the radius of the core, see fig.~\ref{fig:prem variation}.  In our implementation of these formulas, we approximate the integrals by assuming a constant density over intervals of the Earth's radius that are roughly $30\,\text{km}$ wide.

In addition, one could separately constrain the moment of inertia of the Earth, which is measured separately from its mass \cite{1994AJ....108..711W}, but our parameterization allows for that to vary.

\subsection{The Matter Effect}
The matter effect contributes a potential to electron neutrinos in the flavor basis \cite{Wolfenstein:1977ue}, which modifies the propagation of neutrinos in dense media, such as the Earth, the Sun, and in supernova.
The matter effect has been detected in the Sun at modest significance \cite{Gonzalez-Garcia:2013usa} by combining solar neutrino data \cite{SNO:2002tuh} with long-baseline reactor neutrino data \cite{KamLAND:2013rgu}, but only hints exist for the Earth\footnote{Some hints at $<3\sigma$ of the terrestrial matter effect exists in solar neutrinos \cite{Super-Kamiokande:2013mie,Super-Kamiokande:2016yck,yasuhiro_nakajima_2020_4134680} with expected improvements from long-baseline accelerator and solar neutrinos at DUNE \cite{Kelly:2018kmb,Capozzi:2018dat}.}.
Since the neutron fraction differs considerably between the Sun and the Earth ($\sim0.3$ and $\sim1.05$) new physics scenarios such as non-standard neutrino interactions \cite{Wolfenstein:1977ue,Proceedings:2019qno} would manifest differently.
In addition, there is some uncertainty about the chemical composition of the core, leading to a slightly different neutron fraction which could also be probed with oscillations \cite{Rott:2015kwa,Bourret:2017tkw} since, for a fixed density, the neutron fraction affects the electron density.

For simplicity and comparison with existing literature, we parameterize deviations from the expected matter potential within the non-standard neutrino interaction (NSI) framework \cite{Wolfenstein:1977ue,Proceedings:2019qno}.
That is, we rescale the matter effect in the Earth by $a\to a(1+\epsilon_{ee}^\oplus)$ where the $ee$ indicates that this modification is only to electron neutrinos and the $\oplus$ indicates that this is for Earth neutron fractions.

\subsection{Atmospheric Neutrino Detection at DUNE}
DUNE is anticipated to have sensitivity to atmospheric neutrinos down to $E\sim100$ MeV with modest angular reconstruction capabilities \cite{Kelly:2019itm,DUNE:2020ypp}, while Super-KamiokaNDE's (SK) sensitivity falls off below 1 GeV \cite{Super-Kamiokande:2019gzr}, and IceCube's sensitivity falls off below 6 GeV \cite{IceCube:2017lak}.
JUNO may also have sensitivity to low-energy atmospheric neutrinos \cite{JUNO:2021tll}, however they are not expected to have significant angular information, limiting their ability to probe oscillations in the core.
Thus, studying atmospheric neutrinos at DUNE is particularly interesting, as there are unique oscillation effects present below 1 GeV related to the core of the Earth.

The primary detection channels will be charged-current interactions with no pions and either zero or one proton with the former being antineutrino rich and the latter being neutrino rich.
This allows one, in principle, to differentiate among the four detectable neutrinos: $\nu_e$, $\nu_\mu$, $\bar\nu_e$, and $\bar\nu_\mu$.
We assume flavor discrimination but, conservatively, we assume no neutrino/antineutrino discrimination\footnote{When we rescale our event rates to match those in \cite{Kelly:2019itm}, we find a comparable sensitivity to $\delta$.
For our analyses we use our higher event rates which are closer to those described in \cite{DUNE:2020ypp}.}.
In addition, we don't include partially contained $\nu_\mu$ events reducing the $\nu_\mu$ event rate by 25\%, consistent with \cite{DUNE:2020ypp}.

To compute the oscillation probabilities through the Earth, we use \texttt{nuSquids} \cite{ArguellesDelgado:2014rca} and then calculate the event rate in 10 uniformly spaced $\cos\theta_z$ bins from $[-1,1]$ and 43 logarithmically space energy bins from 100 MeV to 8 GeV, which is equivalent to $\sim10\%$ energy resolution.
For the cross section, we use the total $\nu$-$N$ and $\bar\nu$-$N$ cross sections \cite{Formaggio:2012cpf} and conservatively use the $\nu_\mu$ cross section for $\nu_e$ as well.
We then construct a $\Delta\chi^2$ test statistic using poisson log likelihood ratios in the usual way by minimizing over four separate flux normalizations and the spectral index\footnote{While the oscillation parameters affect the flux, we have confirmed that the relative impact for existing constraints on the oscillation parameters is small; with improvements from e.g.~JUNO and DUNE's accelerator neutrinos, the impact will be even smaller. In addition, the mass ordering will be determined with a small amount of exposure at DUNE so we assume the normal mass ordering.}.
Finally, we assume 400 kton-years of exposure, equivalent to 10 years of DUNE's full detector.

\section{Results}
\subsection{The Earth's Core}
\begin{figure}
\centering
\includegraphics[width=\columnwidth]{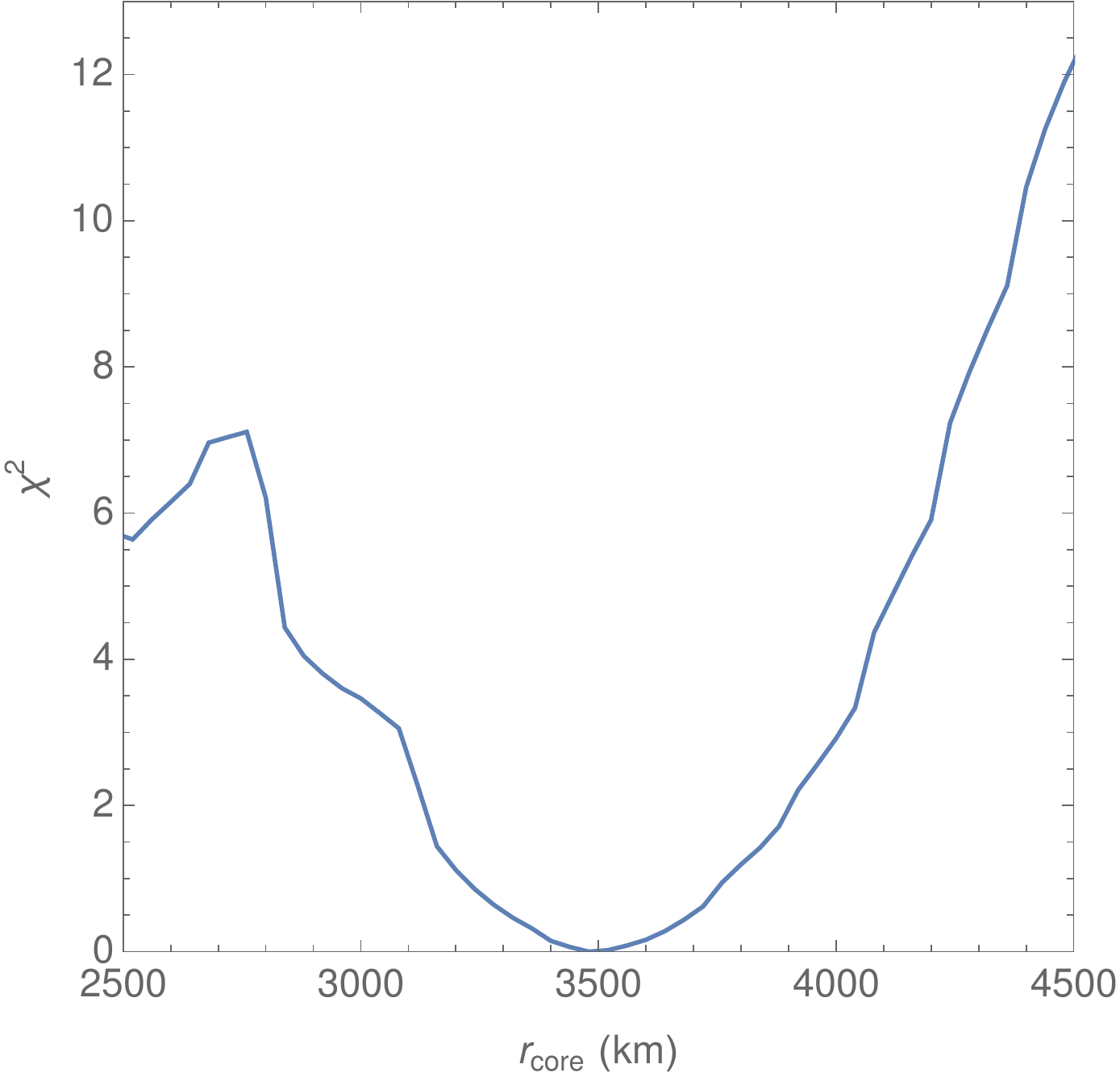}
\caption{The sensitivity to measuring the radius of the Earth's core given 10 years of exposure.}
\label{fig:core}
\end{figure}

We calculated the test statistic comparing the expected Earth model and one where we vary the radius and core of the Earth simultaneously while minimizing over the four flux uncertainties and the spectral index.
Our results are shown in fig.~\ref{fig:core} which shows a strong capability to detect the existence of the core (the test statistic only increases at smaller radii) at $\Delta\chi^2=15$ or $3.9\sigma$.
We also find that DUNE can determine the radius of the core to $\sim 9\%$ precision ($\sim300$ km which is comparable to the size of the D'' layer between the mantle and the core) at $1\sigma$.

\subsection{The Matter Effect in the Earth}
\begin{figure}
\centering
\includegraphics[width=\columnwidth]{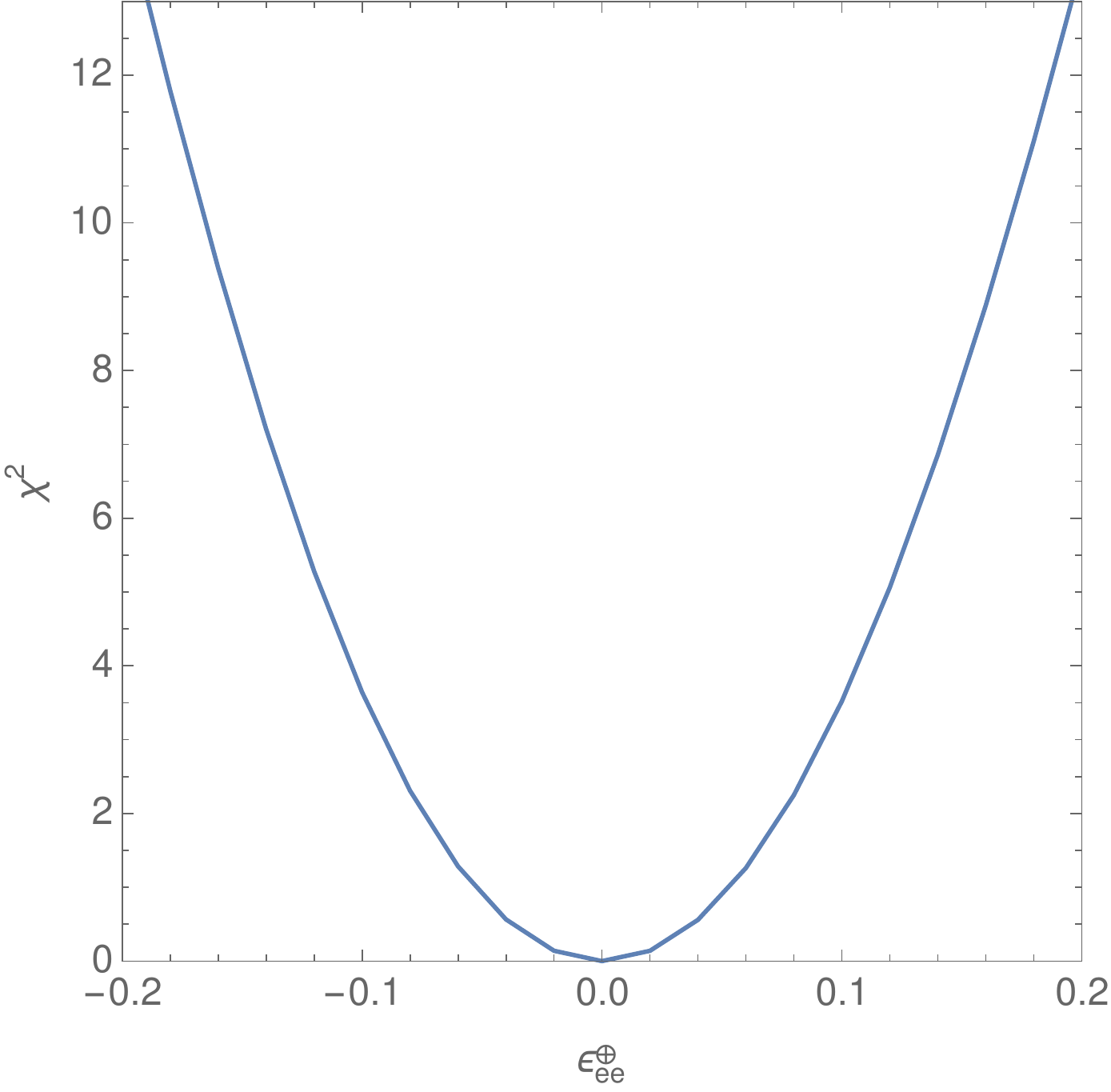}
\caption{The sensitivity to measuring the matter effect in the Earth parameterized by the NSI parameter $\epsilon_{ee}^\oplus$ given 10 years of exposure.}
\label{fig:matter}
\end{figure}

We also investigated whether or not DUNE can determine the existence of the matter effect in the Earth - something that has been only very weakly probed thus far.
We have found that DUNE can determine the matter potential of the Earth at the 5\% level as shown in fig.~\ref{fig:matter}.

For comparison, the best existing evidence for the matter effect in the Earth is from night time solar neutrinos at SK at the $1.9\sigma$ level \cite{yasuhiro_nakajima_2020_4134680}.
Meanwhile, the matter effect in the Sun has been somewhat poorly constrained to $\epsilon_{ee}^\odot\underset{\sim}{\in}[0,1]$ \cite{Esteban:2018ppq,Coloma:2017egw} with the offset due to the slight tension in data from solar and reactor experiments.
In the future, DUNE can determine the existence of the matter effect in the Earth's crust with long-baseline accelerator neutrinos at the $\sim3\sigma$ level, e.g.~$\epsilon_{ee}^\oplus\underset{\sim}{\in}[-0.33,0.33]$.
Thus, a measurement of low-energy atmospheric neutrinos will provide a state-of-the-art probe of the Earth's matter effect.

\section{Conclusions and Future Investigations}
Low-energy atmospheric neutrinos provide a unique probe of the features of the Earth due to rich oscillation phenomenology and DUNE is expected to be sensitive to such oscillations in coming years.
We have shown that oscillations in the Earth are sensitive to the size and density of the Earth's core at the 9\% level at $1\sigma$ and can confirm the existence of the Earth's core at $>3\sigma$.
We have also shown that, assuming we know the size and density of the Earth's core, DUNE can determine the existence of the matter effect in the Earth at the 5\% level at $1\sigma$, equivalent to constraining $\epsilon_{ee}^\oplus\in[0.05,0.05]$ in the non-standard neutrino interaction framework.

We have also shown that, while MSW and parametric resonances could play a role in low-energy atmospheric neutrino oscillations, they are either below thresholds or too small of features to be detected in even a LArTPC detector.

In addition, in the future, it may be possible to combine measurements of the density profile of the Earth from atmospheric neutrino oscillations at DUNE as in this paper, atmospheric oscillations at Super-KamiokaNDE/Hyper-KamiokaNDE \cite{Super-Kamiokande:2019gzr,Hyper-Kamiokande:2018ofw}, nighttime solar neutrinos at SK/HK and DUNE, as well as absorption of atmospheric neutrinos in the Earth at higher energies at JUNO \cite{JUNO:2021tll}, IceCube \cite{Donini:2018tsg,IceCube:2017lak}, KM3NeT \cite{KM3Net:2016zxf}, and Baikal-GVD \cite{BAIKAL:2013jko} to form a 3D constraint of the core of the Earth.

Finally, we note that this study is not exactly probing the size of the Earth's core defined inertially, as is the case for seismography, but rather to the size of the core in terms of the non-universal lepton flavor number (electron number).
We note that this is not only different from seismography studies, but is also different than in \cite{Donini:2018tsg}, which is sensitive to the total weak charge, since most of the absorption in the Earth is due to deep inelastic scattering off quarks.
In principle, if dark matter is captured in the Earth  \cite{Krauss:1985aaa,Gould:1987ir,Green:2018qwo} and has a neutrino interaction \cite{Mori:1992yq,Edsjo:1995zc,Delaunay:2008pc,Lee:2013iua,Lin:2014hla,Xu:2018fac,Reno:2021cdh,Saveliev:2021jtw,Anchordoqui:2018ucj}, this could modify the oscillation probability\footnote{The interaction would need to be flavor non-universal to affect oscillations.}; moreover, if the mediator is lighter than $\mathcal O($few$)$ GeV, then the effect would not show up in absorption tests such as \cite{Donini:2018tsg} but would \emph{only} be seen in an oscillation probe, since the matter effect is a forward elastic scattering process.

\begin{acknowledgments}
We acknowledge helpful discussions with Julia Gehrlein and Pedro Machado as well as support from the US Department of Energy under Grant Contract DE-SC0012704.
The work presented here that RP did was supported in part by the U.S.~Department of Energy, Office of Science, Office of Workforce Development for Teachers and Scientists, Office of Science Graduate Student Research (SCGSR) program. The SCGSR program is administered by the Oak Ridge Institute for Science and Education (ORISE) for the DOE. ORISE is managed by ORAU under contract number DE-SC0014664. All opinions expressed in this paper are the authors' and do not necessarily reflect the policies and views of DOE, ORAU, or ORISE. RP was also supported by the US Department of Energy under Grant Contracts DE-SC0020262 and DE-SC00018327.
Some of the figures and computations were done with \texttt{python} \cite{10.5555/1593511} and \texttt{matplotlib} \cite{Hunter:2007}.
\end{acknowledgments}

\appendix

\section{Resonances}
\label{sec:resonances}
In this appendix we clarify several points on resonant oscillation effects of low-energy atmospheric neutrinos through the Earth and demonstrate the effects numerically for standard oscillation scenarios.

\subsection{MSW Resonance}
Neutrinos in matter are subject to various types of modifications to the standard oscillation probabilities.
The most well known one is the MSW resonance \cite{Mikheyev:1985zog,Wolfenstein:1977ue}.
The term ``MSW'' actually often refers to any one of three separate effects.
First, there is the matter effect \cite{Wolfenstein:1977ue} which modifies the $\Delta m^2$'s, mixing angles, and complex phase in matter and is sometimes erroneously referred to as the MSW effect.
This effect, for example, will play a major role in DUNE's sensitivity to the mass ordering.
Second, there is the adiabaticity effect by which a mixing angle in matter is propagated to vacuum by a sufficiently slowly \cite{Parke:1986jy} varying matter potential.
This is the effect that ultimately solves the solar neutrino problem.
Finally, there is the resonant enhancement of a small mixing angle to a large mixing angle at a certain energy and density which actually appeared earlier in \cite{Wolfenstein:1977ue,Barger:1980tf}.
An example of when this effect is phenomenologically relevant is for $\sim1$ eV sterile neutrino searches at IceCube \cite{Chizhov:1998ug,IceCube:2020phf}\footnote{Should the hints of a sterile neutrino at $\sim1$ eV \cite{LSND:2001aii,MiniBooNE:2018esg,Mention:2011rk,Giunti:2010zu,Barinov:2021asz} be confirmed and its properties measured, the properties of the Earth's core could also be measured with high energy atmospheric neutrinos at IceCube.}.
We now investigate these effects in the present context.

First, neutrinos do experience the regular matter effect which leads to a modification of the oscillation parameters with a change in each layer and the largest change in the core due to its large density.

Second, the adiabaticity effect does not apply here since atmospheric neutrinos are produced very close to vacuum and then experience fairly sharp changes in density as they enter the Earth and move through each of its layer.

Third, there could be a resonant enhancement of small mixing angles into larger effective angles in matter.
Such a resonance, for solar parameters with neutrinos, appears at
\begin{equation}
2\sqrt2G_FN_eE=\Delta m^2_{21}\frac{\cos2\theta_{12}}{c_{13}^2}\,.
\end{equation}
In the crust, mantle, and core, this corresponds to energies of $\sim130$ MeV, $\sim70$ MeV, and $\sim35$ MeV, respectively and can be confirmed by the expressions for $\theta_{12}$ in matter \cite{Denton:2016wmg,Denton:2018hal}.
The effect in the mantle and the core are at too low of energy to observe, and only a very small fraction of the flux passes only through the crust, $\cos\theta_z\in[-0.1,1]$ on land and $[-0.05,1]$ for a detector in the ocean.
In addition, the dominant mixing parameter, $\theta_{12}\sim33^\circ$, is already fairly large -- a further enhancement to $45^\circ$ is only a 20\% increase in the amplitude of the oscillation probability.
Finally, the enhancement only exists for a fairly narrow energy region which must then also line up with an oscillation maximum which does not happen in the crust.
This is because the first maximum at $E=130$ MeV is at $\cos\theta_z\sim-0.2$ -- well into the mantle.
Thus there are no MSW resonances for atmospheric neutrinos at DUNE.

\subsection{Parametric Resonance}
Another type of resonance is known as a parametric resonance which can also amplify a small mixing angle into a large one \emph{even in the presence of a modest average matter potential}, provided that the potential has a certain shape, the so-called castle-wall potential \cite{Akhmedov:1988kd,Krastev:1989ix}.
In this case, for tuned conditions, neutrinos experience a half integer number of oscillations in one part of the potential, and then another half integer number of oscillations in the next part, and so on, leading to a resonant enhancement of the total oscillation.
The closest environment to a castle-wall potential in nature is the Earth's mantle and core where neutrinos go from a modest potential in the mantle, to a high potential in the core, and then back in a modest potential in the mantle again.
If the neutrino energy and path lengths are just right, then a considerable increase in the mixing can occur.

More precisely, this occurs when the accumulated phase in each segment is an odd integer multiple of $\pi/2$.
That is, when the following two equations are simultaneously satisfied in the mantle and the core for some integers $k_m,k_c$,
\begin{align}
\left|\frac{\Delta\wt{m^2_{21}}_mL_m}{4E}\right|&=\frac\pi2(2k_m+1)\,,\\
\left|\frac{\Delta\wt{m^2_{21}}_cL_c}{4E}\right|&=\frac\pi2(2k_c+1)\,,
\end{align}
where the effective $\Delta\wt{m^2_{21}}$ in matter (see e.g.~\cite{Denton:2019yiw}) is well described by
\begin{equation}
\Delta\wt{m^2_{21}}=\Delta m^2_{21}\sqrt{(\cos2\theta_{12}-c_{13}^2a/\Delta m^2_{21})^2+\sin^22\theta_{12}}\,,
\end{equation}
where the density in the mantle or core appears in the matter effect, $a$.
The propagation distances are,
\begin{align}
L_m&=-r_E\cos\theta_z-\sqrt{r_c^2-r_E^2\sin^2\theta_z}\,,\\
L_c&=2\sqrt{r_c^2-r_E^2\sin^2\theta_z}\,.
\end{align}

\begin{figure*}
\centering
\includegraphics[width=0.49\textwidth]{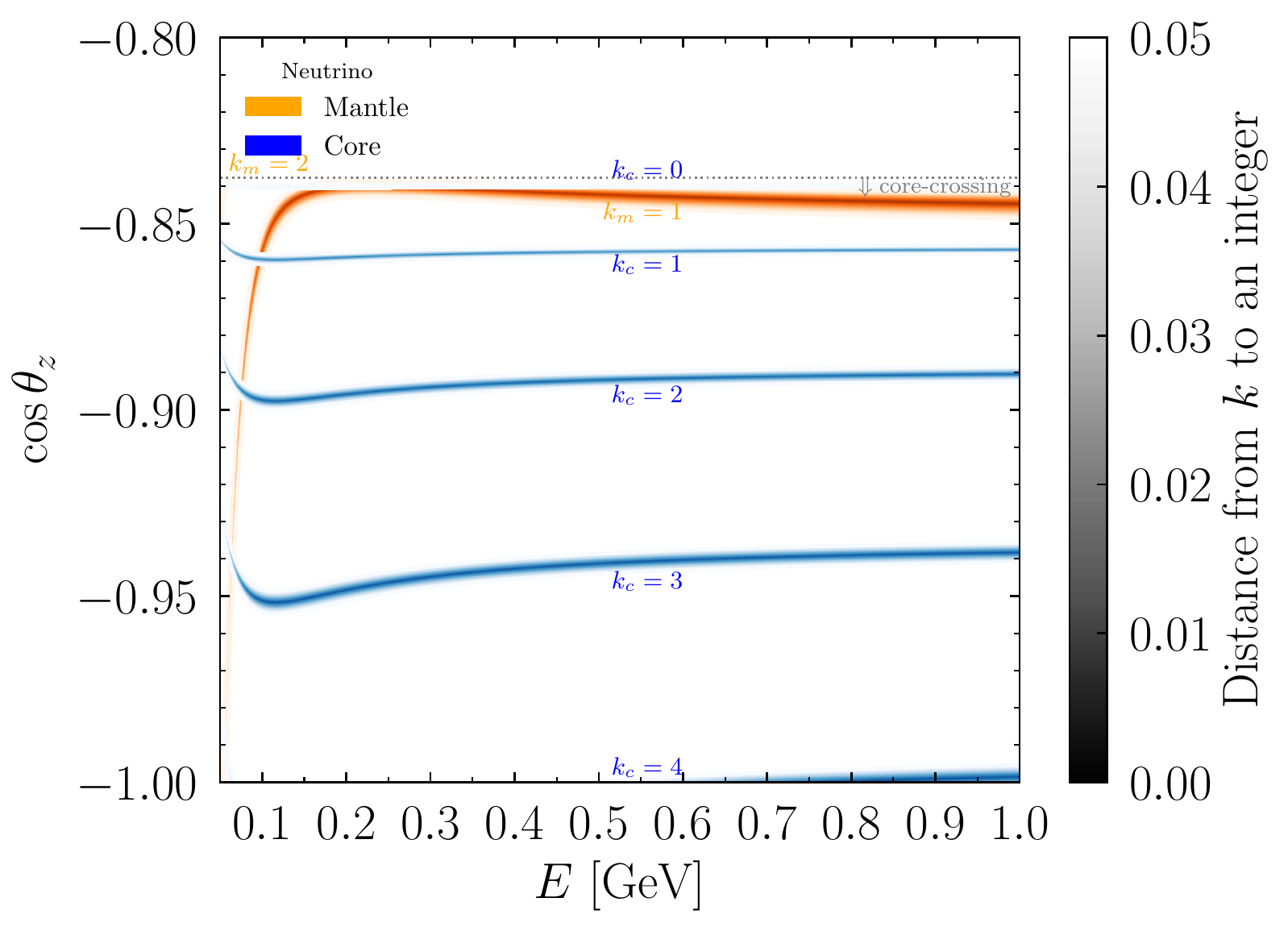}
\includegraphics[width=0.49\textwidth]{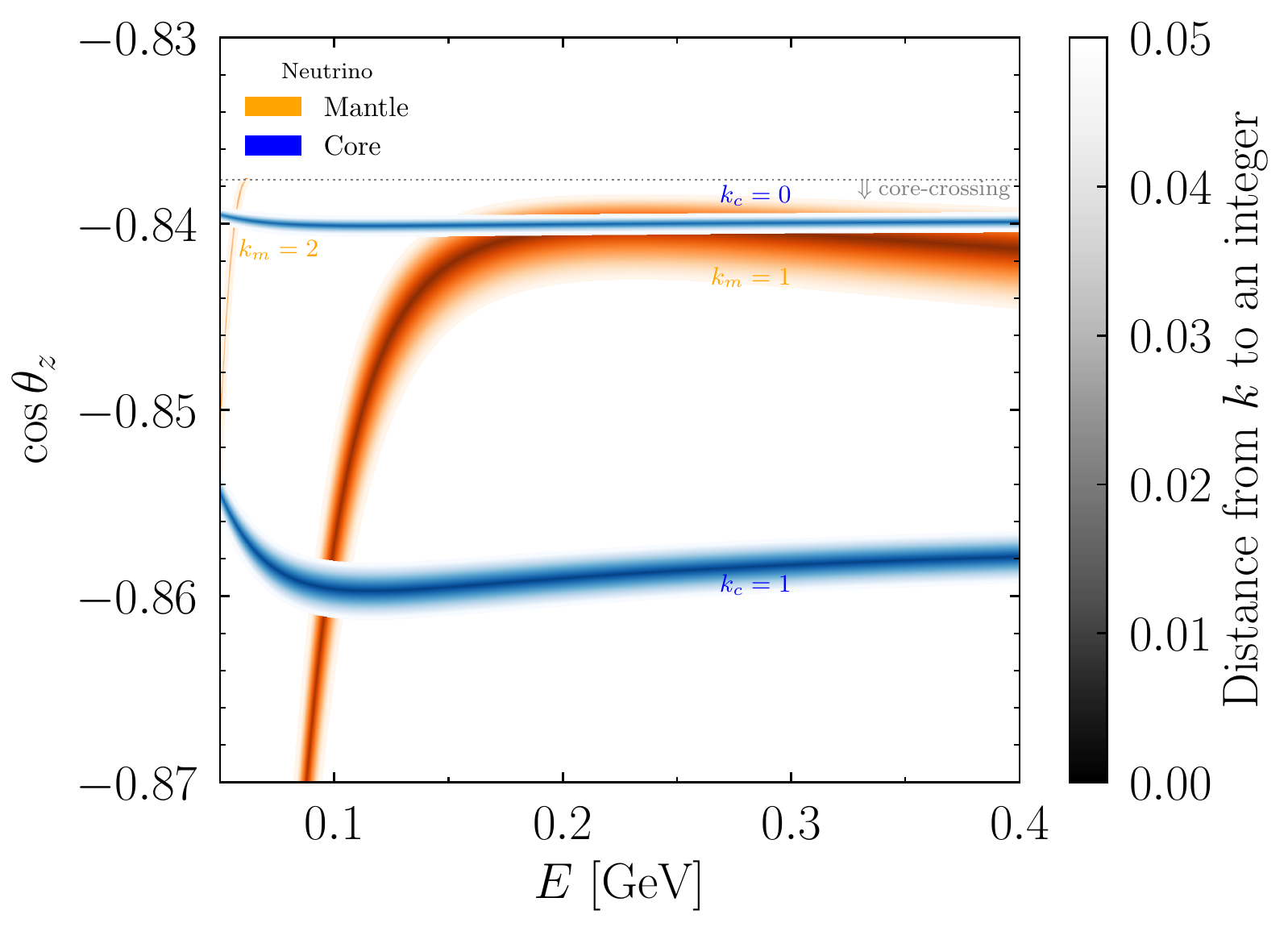}
\includegraphics[width=0.49\textwidth]{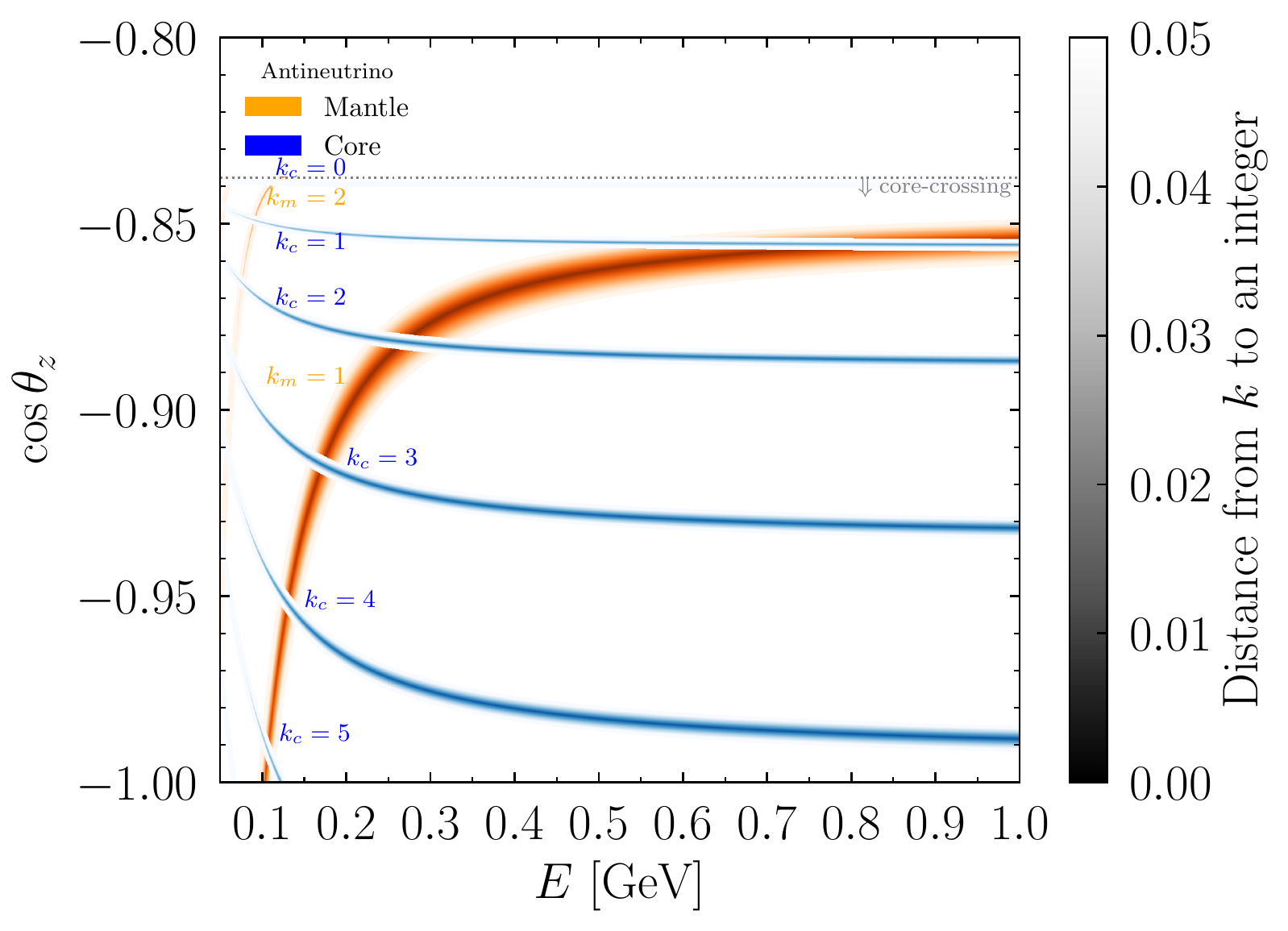}
\includegraphics[width=0.49\textwidth]{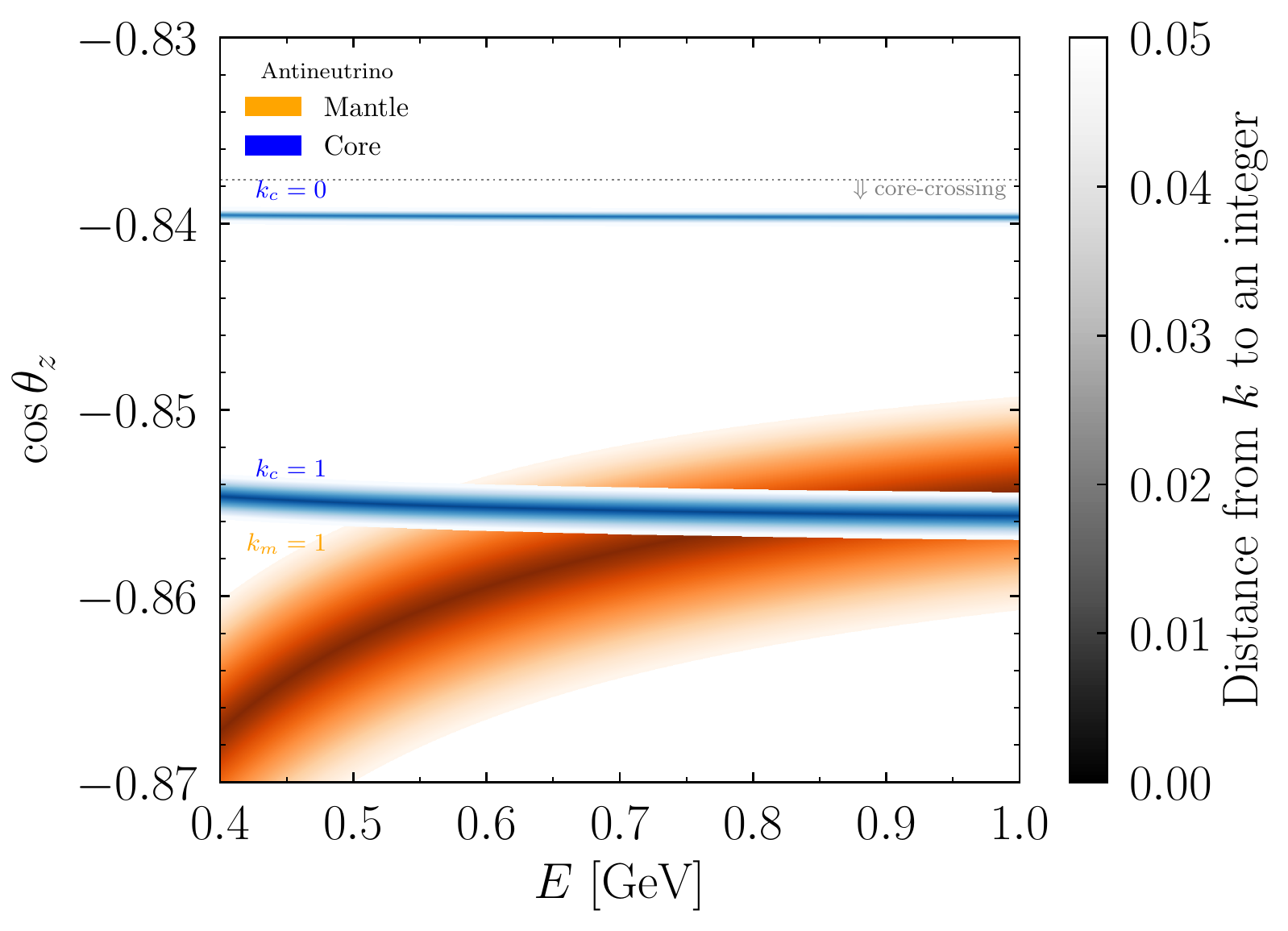}
\caption{Here we plot the distance that $k_c,k_m$ is from an integer where $m,c$ refers to the mantle (orange) and core (blue) respectively, as a function of neutrino energy and $\cos\theta_z$.
Darker colors are closer to integers and thus closer to a parametric resonance.
A parametric resonance is only achieved when both $k_c$ and $k_m$ are close to integers; these are the overlap regions in these plots.
The top panels are for neutrinos and the bottom for antineutrinos, the right panels are zoomed in on the largest overlap regions.
Note that the energy scale of some panels goes slightly below DUNE's anticipated threshold.}
\label{fig:parametric}
\end{figure*}

\begin{figure}
\centering
\includegraphics[width=\columnwidth]{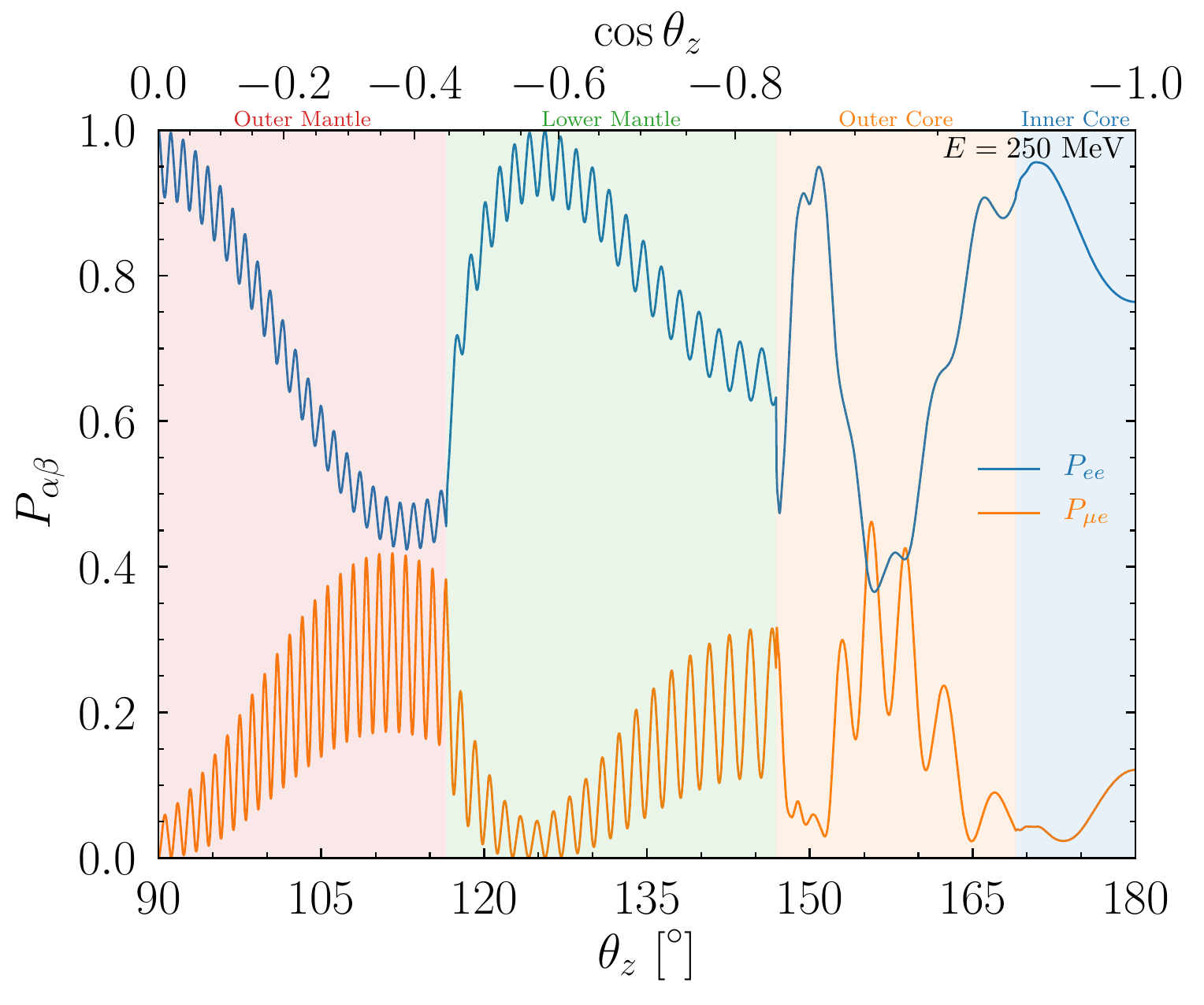}
\caption{The $\nu_e\to\nu_e$ (blue) and $\nu_\mu\to\nu_e$ (orange) probabilities as a function of $\theta_z$ and $\cos\theta_z$ for $E=250$ MeV propagated through a four layer Earth model.}
\label{fig:P cosz}
\end{figure}

We then show numerically, in fig.~\ref{fig:parametric}, the locations of the different resonances for neutrinos and antineutrinos by plotting how close $k_m$ and $k_c$ are to integers.
For these figures we assume that the Earth is composed of only two layers, a mantle at $\rho=5.5$ g/cc and a core at $\rho=11$ g/cc with $Y_e=\frac12$ and $r_c=3480$ km.
We found that there is an extremely narrow region of parameter space where a parametric resonance does appear; that is, both $k_m$ and $k_c$ are near integers in similar regions of parameter space.
This region spans a modest range in energy from $\sim175$ MeV to $\sim325$ MeV, but only a very narrow angular range\footnote{The fact that the largest region of parametric resonance in the Earth is near the core-mantle boundary is largely a coincidence as can be seen by the smooth transition of the $k_m\simeq1$ region to smaller values of $\cos\theta_z$.} around $\cos\theta_z\sim-0.84$ with $\Delta\cos\theta_z\lesssim0.001$.
For antineutrinos we see a similar effect at higher energies ($\sim700$ MeV to $\sim1$ GeV) and slightly longer path lengths ($\cos\theta_z\sim-0.855$ and $\Delta\cos\theta_z\sim0.002$).
These regions are at $k_m=1$ and $k_c=0$ which means that neutrinos experience one and a half solar wavelengths in the mantle, then a half a solar wavelength in the core, then one and a half solar wavelength in the mantle again.
The neutrinos experience many atmospheric wavelengths, the effects of which will be smeared out.
For antineutrinos it is for $k_m=k_c=1$, thus there is an extra full solar oscillation in the core in order for the resonance conditions to be met.
Further resonances exist for more oscillations in the core, but only at energies below 100 MeV, and only for exceedingly narrow regions in parameter space in both $\cos\theta_z$ and energy.

For both neutrinos and antineutrinos the angular width of the parametric resonance is much narrower than any conceivable angular precision ($\Delta\cos\theta_z=0.001$ maps onto $\Delta\theta\sim0.1^\circ$ for these values of $\cos\theta_z$).
Moreover, in between integer values of $k_c$ are half-integer values where the opposite effect occurs: the partial enhancement that may happen in the mantle is damped in the core (this would occur, e.g.~on an orange curve and halfway between blue curves in fig.~\ref{fig:parametric}).
Thus smearing across an experimentally realistic range of $\cos\theta_z$ completely removes the enhancement.
Since the successive integer $k_c$ values occur at a spacing of $\Delta\cos\theta_z\sim0.05$ which corresponds roughly to $\sim5^\circ-10^\circ$, any experiment measuring neutrinos at these energies will only observe an average of multiple parametric enhancements and suppressions.
As such, we return to the usual matter effect which modifies the mixing angles in the usual way \cite{Wolfenstein:1977ue,Barger:1980tf,Zaglauer:1988gz,Kimura:2002wd,Denton:2019ovn}.

Finally, the small size of the parametric effects are confirmed in fig.~\ref{fig:P cosz} which shows the appearance and disappearance neutrino oscillation probabilities at a fixed energy, 250 MeV, as a function of zenith angle propagated through four layers of the Earth in a full three flavor picture.
We see that, while there is a resonant feature that appears at $\cos\theta_z\sim-0.84$, it is extremely narrow and don't reach beyond the amplitude of normal oscillations in matter through the mantle anyway.
The remaining features after that are generally oscillatory and are not related to a resonant effect, and will also smear out as well.

\bibliography{main}

\end{document}